\documentclass[11pt,twoside]{article}
\usepackage{asp2010}

\resetcounters

\begin{document}

\title{CIGALE: Code Investigating GALaxy Emission}
\author{Yannick Roehlly$^1$, Denis Burgarella$^1$, V\'eronique Buat$^1$,
\'Elodie Giovannoli$^1$, Stefan Noll$^2$, and Paolo Serra$^3$
\affil{$^1$Laboratoire d'Astrophysique de Marseille / C.N.R.S. - Universit\'e de
Provence\\
38, rue Fr\'ed\'eric Joliot-Curie, 13388 Marseille, France}
\affil{$^2$Institut f\"ur Astro- und Teilchenphysik, Universit\"at Innsbruck,\\
Technikerstr. 25/8, 6020 Innsbruck, Austria}
\affil{$^3$Astrophysics Branch, NASA/Ames Research Center,\\ MS 245-6, Moffett Field, CA 94035}}

\begin{abstract}
We present CIGALE \citep{2005MNRAS.360.1413B,2009A&A...507.1793N}, a software developed at the \textit{Laboratoire d'Astrophysique de Marseille} to fit galaxy spectral energy distributions from the rest-frame far-UV to far-IR wavelength range, and to derive some of their physical parameters. We also give some examples of scientific results obtained with CIGALE.
\end{abstract}

\section{Aim of the software}

The multi-wavelength observation of galaxies allows astrophysicists to derive some of their physical parameters from the comparison of their spectral energy distributions (SEDs) to computed SEDs based on models and templates (SED fitting). Scientists from the \textit{Laboratoire d'Astrophysique de Marseille}\footnote{\url{http://lam.oamp.fr}} developed CIGALE (/si.gal/), a Code Investigating Galaxy Emission, that takes into account both the dust ultraviolet-optical attenuation and its corresponding infra-red re-emission. CIGALE is able to statistically derive reliable physical parameters from UV to IR observations.

\section{How CIGALE works}

\begin{figure}[htp]
 \centering
 \includegraphics{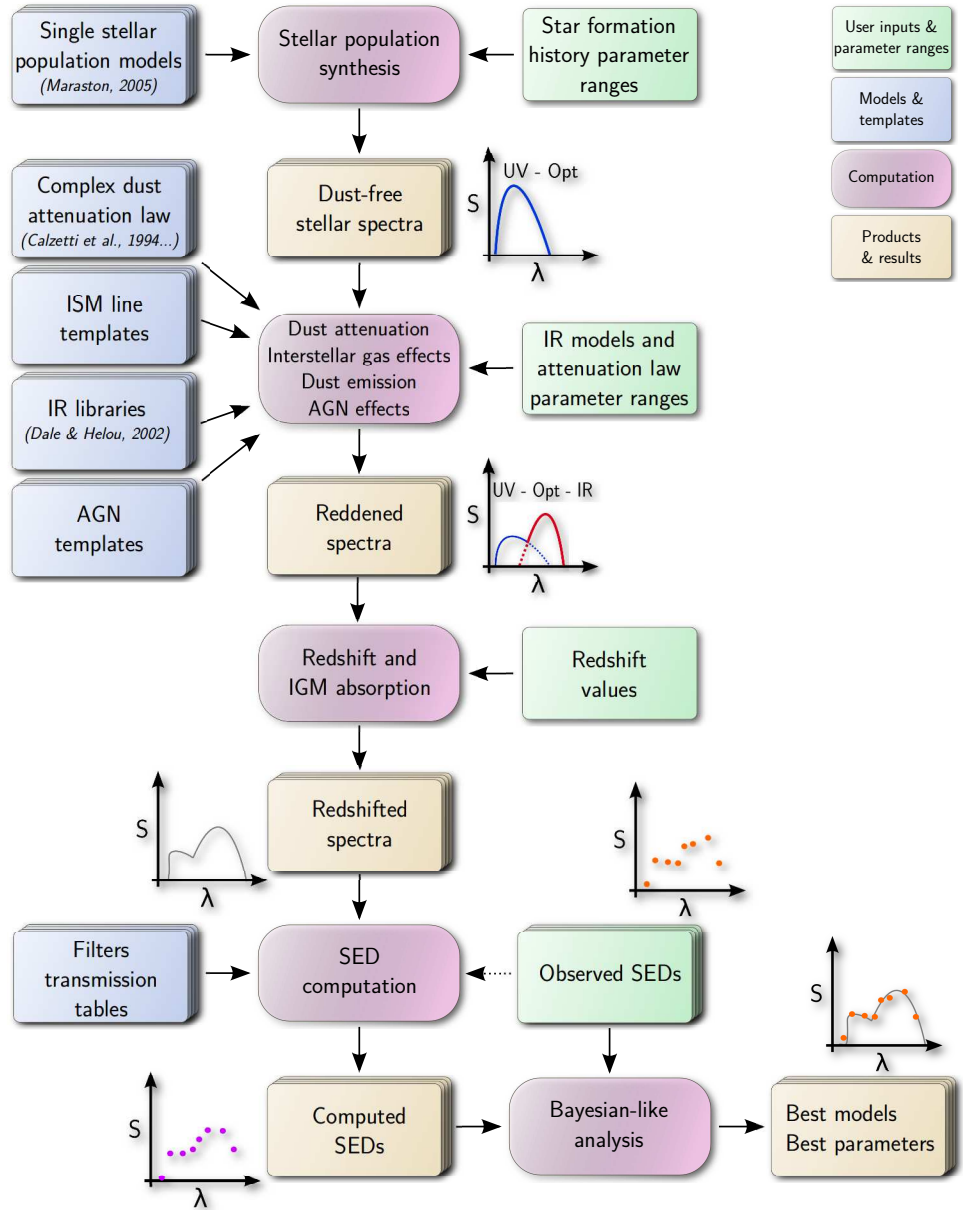}
 \caption{CIGALE operation workflow}
 \label{fig:workflow}
\end{figure}

The user provides CIGALE with multi-$\lambda$ fluxes and redshift for each galaxy, and with a list of possible values for each physical parameter. The parameters are those related to star formation history ($\tau$ and ages for young and old stellar populations (SP), mass fraction of young SP), dust attenuation ($V$-band attenuation, reduction factor of $A_{v}$ for old SP) and dust emission (IR power-law slope, AGN related fraction of $L_{dust}$).

\begin{figure}[htp]
 \centering
 \includegraphics{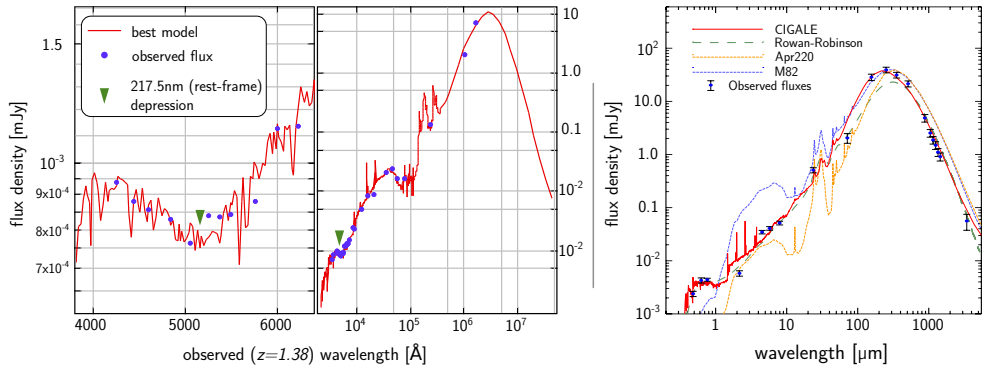}
 \caption{Examples of scientific use. Left: $z=1.38$ galaxy SED showing $217.5nm$ depression
 \citep{2011A&A...533A..93B}. Right: CIGALE fitting of HLSW-01 - adapted from \citet{2011ApJ...732L..35C}.}
 \label{fig:scientific_use}
\end{figure}

Using various models, libraries and templates (see figure~\ref{fig:workflow}) CIGALE computes all the possible spectra and derives mean fluxes in the observed filter bands. Then a Bayesian-like statistical analysis permits to determine for each galaxy the best value for each parameter as well as the best computed model (see figure~\ref{fig:scientific_use}).

\section{Example of scientific applications}

\subsection{High-redshift galaxy study}

\citet{2011A&A...533A..93B} combined photometric data with broad and intermediate band filters, and far-infrared  data from Herschel to sample the ultraviolet spectrum of  high redshift galaxies and characterize their dust extinction (see figure~\ref{fig:scientific_use} left). A depression at $217.5nm$ is clearly identified in all sources.  An analytical formula of the dust attenuation curve is deduced from CIGALE with a bump  of moderate amplitude at $217.5nm$ \citep[see also][]{2011ApJ...734L..12B,2011A&A...525A.150G}.

\subsection{Multiply-lensed galaxy study}

\citet{2011ApJ...732L..35C} used CIGALE to study the multiply-lensed galaxy HLSW-01 and compute some of its physical properties, in particular its stellar mass, its star formation rate and the proportion of young stars. The right part of figure~\ref{fig:scientific_use} compares the best CIGALE model to the best combination of templates from \citet{2010MNRAS.409....2R} and IR SEDs of Arp220 and M82.

\section{CIGALEMC - Using Monte Carlo Markov Chain statistical method}

From the CIGALE code base \citet{2011ApJ...740...22S} developed CIGALEMC that uses a Monte Carlo Markov Chain method to find the best fit parameters. CIGALEMC is made to be efficient (needed CPU time grows linearly, not exponentially, with the number of fitted parameters), accurate (statistical quantities are robustly determined using Gelman \& Rubin diagnostic as convergence criteria) and user friendly (\textit{a priori} deciding the parameter density, to find a compromise between accuracy and speed, is not necessary).

\section{Future evolutions}

In addition to \citet{2002ApJ...576..159D}, we want to propose the use of \citet{2001ApJ...556..562C}, \citet{2007A&A...461..445S} and \citet{2007ApJ...657..810D} libraries, as well as the use of CB07 stellar population model  from Bruzual and Charlot\footnote{\url{http://bruzual.org/cb07/}} in addition to \citet{2005MNRAS.362..799M}.

We are studying the port of CIGALE to Python -- presently the code is written in Fortran -- for more modularity, more readability of the code and more evolution opportunities.

\section{Contact and download}

CIGALE is available for download on its web site\footnote{\url{http://cigale.oamp.fr}}, where you will also find an on-line, java applet based, version. For more information on CIGALE software, you can contact Denis Burgarella $<$denis.burgarella@oamp.fr$>$.

\bibliographystyle{asp2010}
\bibliography{P131}

\end{document}